\documentclass[]{tPHM2e}
\usepackage{graphicx}
\newcommand{\YRS}{YbRh$_2$Si$_2$}
\newcommand{\YRSs}{YbRh$_2$Si$_2$ }
\newcommand{\YIS}{YbIr$_2$Si$_2$}
\newcommand{\YISs}{YbIr$_2$Si$_2$ }

\newcommand{\YTS}{YbT$_2$Si$_2$}
\newcommand{\TCS}{ThCr$_2$Si$_2$}
\newcommand{\CBG}{CaBe$_2$Ge$_2$}

\newcommand{\YRIS}{Yb(Rh$_{1-x_{\rm Ir}}$Ir$_{x_{\rm Ir}})_2$Si$_2$}
\newcommand{\YRCS}{Yb(Rh$_{1-x_{\rm Co}}$Co$_{x_{\rm Co}})_2$Si$_2$}

\newcommand{\EA}{\textit{et al.}}

\begin{document}
\doi{10.1080/14786435.20xx.xxxxxx}
\issn{1478-6443}
\issnp{1478-6435}
\jvol{00} \jnum{00} \jyear{2010} 

\markboth{C. Krellner et al.}{Single crystal growth of YbRh$_2$Si$_2$ and YbIr$_2$Si$_2$}

\articletype{Article}

\title{Single crystal growth of YbRh$_2$Si$_2$ and YbIr$_2$Si$_2$}

\author{Cornelius Krellner$^{\rm a,b}$$^{\dagger}$\thanks{$^\dagger$Corresponding author. Email: krellner@cpfs.mpg.de},
Sebastian Taube$^{\rm a}$$^{\ddagger}$\thanks{$^\ddagger$Present address: Siltronic AG, D-09599 Chemnitz, Germany},
Tanja Westerkamp$^{\rm a}$$^{\S}$\thanks{$^\S$Present address: Deutsches Biomasseforschungszentrum, D-04347 Leipzig, Germany},
Zakir Hossain$^{\rm c}$,
and 
Christoph Geibel$^{\rm a}$\\
$^{\rm a}${\em{Max Planck Institute for Chemical Physics of Solids, D-01187 Dresden, Germany}}\\
$^{\rm b}${\em{Cavendish Laboratory, University of Cambridge, Cambridge CB3 0HE, United Kingdom}}\\
$^{\rm c}${\em{Department of Physics, Indian Institute of Technology, Kanpur 208016, India}}\\
\received{\today}
}

\maketitle

\begin{abstract}
We report on the single crystal growth of the heavy-fermion compounds \YRSs and \YISs using a high-temperature indium-flux technique. The  optimization of the initial composition and the temperature-time profile lead to large (up to 100\,mg) and clean ($\rho_0\approx0.5\,\mu\Omega$cm) single crystals of \YRS. Low-temperature resistivity measurements revealed a sample dependent temperature exponent below 10\,K, which for the samples with highest quality deviates from a linear-in-$T$ behaviour. Furthermore, we grew single crystals of the alloy series \YRIS\, with $0\leq x_{\rm Ir} \leq 0.23$ and report the structural details. For pure \YIS, we establish the formation of two crystallographic modifications, where the magnetic $4f$-electrons have different physical ground states.
\end{abstract}
\bigskip

\bigskip

\section{Introduction}
Most of the ternary intermetallic compounds with the chemical formula AT$_2$X$_2$ ($\rm A=$ alkaline earth metals, lanthanides or actinides, $\rm T=$ transition metal and $\rm X=B$, Ga, Si, Ge, Sn, P, As, ...) crystallize in ternary variants of the tetragonal body-centred BaAl$_4$ structure type, usually in the \TCS-structure one \cite{Ban:1965, Hoffmann:1985}. Characteristic of this structure are layers of edge-connected TX$_4$-tetrahedra, which alternate along the $c$-direction with planar square lattices of A-atoms. Prominent examples, concerning there physical properties, are CeCu$_2$Si$_2$ \cite{Bodak:1966}, URu$_2$Si$_2$ \cite{Hiebl:1983} and SrFe$_2$As$_2$ \cite{Pfisterer:1980}, which all are model systems among the field of strongly correlated electron systems.

Within this field, the discovery of remarkable phases and transitions is often tightly coupled to the growth and characterization of novel materials. Studying these emergent phenomena necessitate the preparation of high-quality single crystals to investigate the physical properties as a function of crystal orientation and momentum. Here, we report on the details of the single crystal growth of pure \YRSs and \YISs as well as the alloy series \YRIS, using a high-temperature indium-flux technique. The growth of sizeable single crystals of these compounds is challenging, because of the high vapour pressure of Yb, whose boiling point, $1196^{\circ}$C, is well below the melting point of Rh ($1964^{\circ}$C) or Ir ($2466^{\circ}$C). Thus, the growth from a stoichiometric melt is likely impossible. Furthermore, the solubility of Si, Rh and Ir in the most appropriate low-melting solvents, In and Sn, is small, making the growth of large crystals rather difficult.

\YRSs also crystallizes in the ThCr$_2$Si$_2$ structure \cite{Rossi:1979} and was intensively studied in recent years, because it is situated on the magnetic side ($T_N=72$ mK) of, but very close to a quantum critical point (QCP). The proximity to the QCP leads to pronounced non-Fermi-liquid behavior in transport and thermodynamic properties, such as the divergence of the electronic Sommerfeld coefficient, $\gamma=C^{4f}/T$, and a linear-in-$T$ resistivity \cite{Trovarelli:2000a}. The weak antiferromagnetic (AFM) order can be continuously suppressed by applying a tiny magnetic field \cite{Gegenwart:2002, Custers:2003}. Electrical transport and thermodynamic measurements have revealed multiple vanishing energy scales \cite{Gegenwart:2005} and, when the finite-temperature results are extrapolated to the $T = 0$ limit, a discontinuity of the Fermi volume across the QCP \cite{Paschen:2004, Friedemann:2010, Hartmann:2010}.

A specific feature of the \YRSs single crystals are their suitability to cleave, yielding atomically flat and clean surfaces perpendicular to the crystallographic $c$-direction. This made extensive angle-resolved photo-emission spectroscopy possible, which gives significant insight into the heavy-fermion behaviour of $4f$-based correlated systems \cite{Danzenbacher:2007, Vyalikh:2008, Vyalikh:2010}. Recent tunneling-spectroscopy experiments succeeded to directly image the transformation into the coherent ground state of the Kondo lattice \cite{Ernst:2011}.

\begin{figure}[t]
\begin{center}
\includegraphics[width=0.8\textwidth]{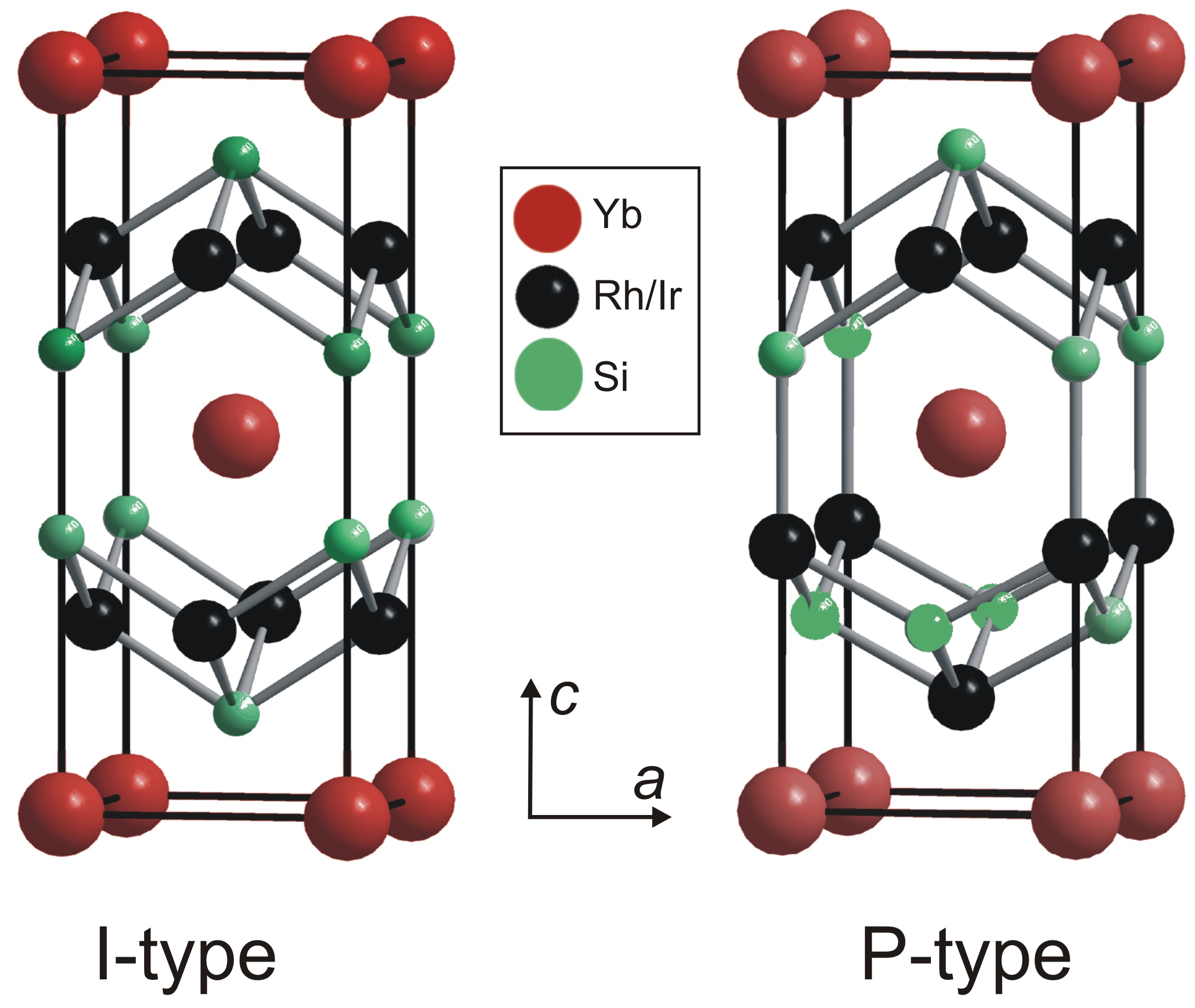}
\caption{\label{fig1} Unit cell of the \TCS\, and the \CBG\, structure types, denoted as I-type and P-type, respectively. \YRSs crystallizes with the I-type structure, whereas, both structures form in the case of \YIS.}%
\end{center}
\end{figure}

The isoelectronic compound \YISs crystallizes in two different structure types, the ThCr$_2$Si$_2$ and the \CBG\, \cite{Hossain:2005}. In the latter, an Ir- and Si-plane are interchanged with each other, therefore, the \CBG\, structure has a lower symmetry, with a missing mirror plane around the body centred Yb-atom. This structure is therefore named primitive (or P-type) in contrast to the body-centred (I-type) version. In Fig.\ref{fig1}, both structure types are presented and one clearly can see the interchanged Ir-/Si-atoms in the lower Ir-Si-plane.

Low-temperature resistivity and specific-heat measurements of I-type \YISs show strong similarities with \YRS. However below 200\,mK, I-type \YISs presents a magnetically non-ordered Fermi-liquid ground state and is  situated on the paramagnetic side of the QCP. A larger unit-cell volume, compared to \YRS, results in a higher Kondo-energy scale, leading to a Kondo-screened ground state. Therefore, the series \YRIS\, is well suited to study the physical properties across this QCP. First studies revealed a detaching of the AFM QCP from the Fermi-surface reconstruction and a new spin-liquid-type ground state for $x_{\rm Ir}=0.06$ and $0.17$ \cite{Friedemann:2009a, Friedemann:2010a}. On the other hand, P-type \YISs orders antiferromagnetically at $T_N\approx 0.6$\,K and do not present any non-Fermi-liquid characteristics \cite{Hossain:2006}.

\section{Single crystal growth}
\subsection{Experimental}
The single crystal growth of \YRSs and \YISs was achieved using an In-flux technique \cite{Canfield:2001, Trovarelli:2000a, Hossain:2005}. Yb ingots (Ames, 99.99\%), Rh/Ir powder (Heraeus, 99.95\%, Si pieces (Alfa Aesar, 99.9999\%) and In shots (Alfa Aesar, 99.9999\%) were put together in a conical Al$_2$O$_3$-crucible ($V\approx30$\,ml). The Yb was stored and weighted in a glove box filled with purified argon gas. Great care was undertaken to keep the oxygen and water content below 0.1\,ppm. Afterwards, the Al$_2$O$_3$-crucible was enclosed in a Ta-tube, again under very pure Ar-atmosphere, using arc welding. Subsequently, the Ta-tube was put under Ar in a vertical high-temperature resistance furnace, were temperatures of up to $1600^{\circ}$C could be achieved. The temperature was measured in-situ at the bottom of the crucible, independently of the furnace temperature. For cooling, a modified Bridgman technique was utilized by slowly moving the Ta-tube out of the furnace with velocities as low as 0.1\,mm/h. After the growth, the In-flux was removed by etching the solidified melt in slightly diluted hydrochloric acid at moderate temperatures ($\approx 60^{\circ}$C).  This dissolves also most of the binary and ternary foreign phases which has been formed due to the off-stoichiometric starting composition. The \YTS\, single crystals were found to be very stable in HCl, therefore, a complete etching for 1-2 days was possible. Finally, the single crystals were rinsed in an ultrasonic bath using acetone.

\subsection{\YRS}
The growth of \YRSs single crystals from a tin-flux was already reported in the first publication of its low-temperature properties in 2000 \cite{Trovarelli:2000a}. In the meantime, comprehensive studies were carried out on this model-type quantum-critical material, as discussed in the introduction. Most of these measurements were performed on single crystals, obtained from an improved flux-growth process, however, a  description of the growth parameters was not yet reported. In this section, we discuss in detail the parameters of this high-temperature tin-flux technique. Many of the aforementioned experiments were possible only, because of the larger and purer single crystals obtained after the optimization. 

Two different research groups have also been working on the single crystal growth of \YRS. Knebel \EA\, have grown mono-isotypic $^{174}$\YRSs single crystals from In-flux with a residual resistivity, $\rho_0\approx 1.3\,\mu\Omega$cm, as well as very pure non-mono-isotypic crystals \cite{Knebel:2006}. On the latter, de-Haas-van-Alphen and resistivity measurements under pressure were performed. Hu \EA\, reported on very large single crystals, which could be obtained using Zn-flux \cite{Hu:2007}. However, these crystals present only a residual resistivity, $\rho_0\approx 8\,\mu\Omega$cm, likely resulting from Zn-incorporation and no further physical measurements were published until now.

The optimization of the single crystal growth from a metallic flux focuses mainly on two parameters. Firstly, the  initial stoichiometry and secondly, the temperature-time profile of the crystal growth. These two parameter classes are not independent from each other, therefore, we first have optimized the initial stoichiometry and in a second step determined the best-possible temperature-time profile.

\begin{figure}[t]
\begin{center}
\includegraphics[width=0.8\textwidth]{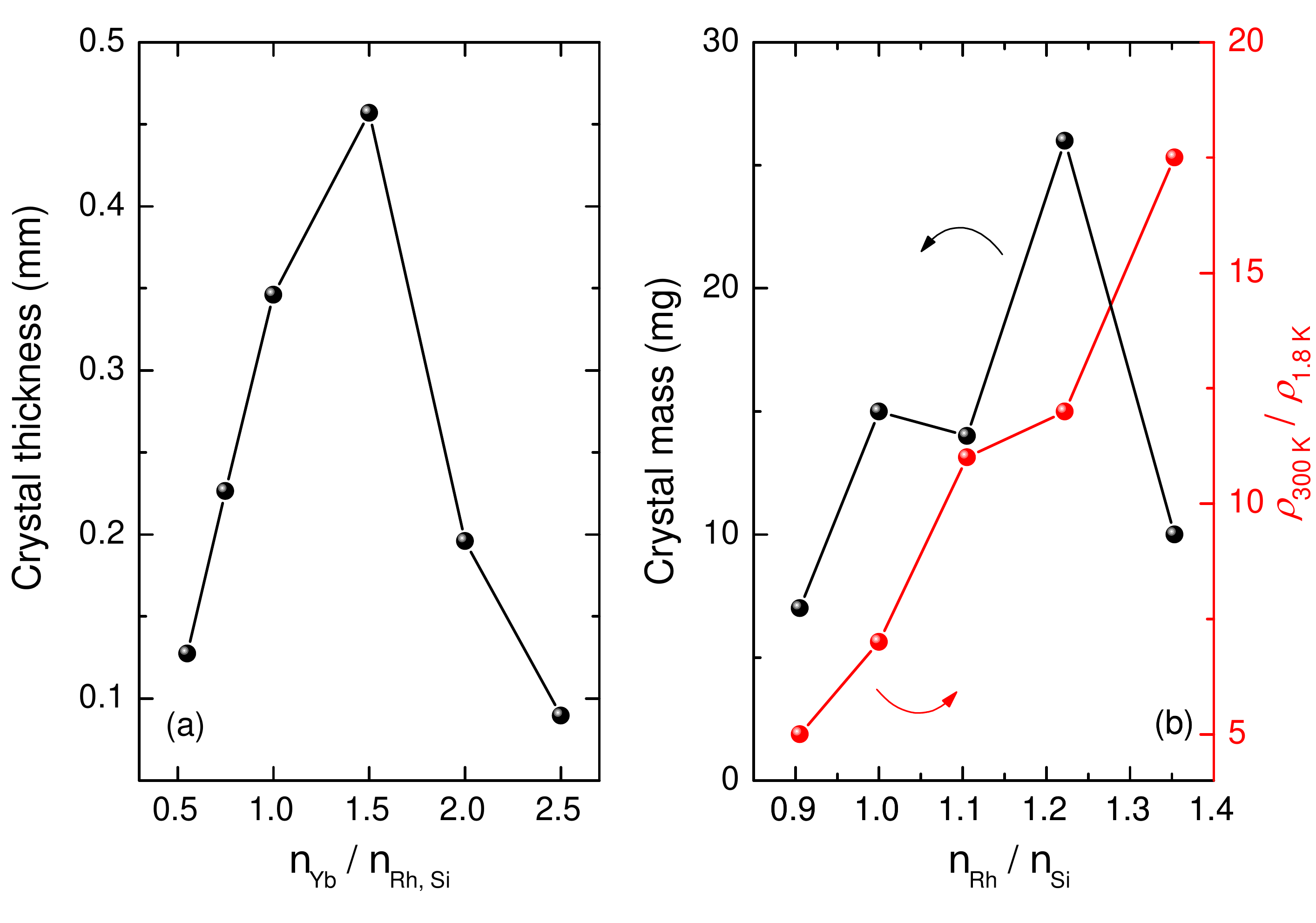}
\caption{\label{ratio} 
Optimization of the initial stoichiometry of \YRS. (a) Influence of the Yb-content on the average crystal thickness. (b) Influence of the Rh-to-Si ratio, $n_{\rm Rh}/n_{\rm Si}$, on the average crystal mass (left axis) and the resistivity ratio, $\rho_{300\rm K}/\rho_{1.8\rm K}$ (right axis).}%
\end{center}
\end{figure}

In the first step, we have varied the initial atomic composition of Yb, Rh and Si with respect to each other, but kept the mass ratio to the amount of In-flux constant, with $m_{\rm In}/m_{\rm Yb,Rh,Si}= 24 M_{\rm In}/M_{\rm Yb,Rh,Si}$. Here, $m_{\rm Yb,Rh,Si}$ is the sum of the mass of Yb, Rh and Si, $M_{\rm In}$ is the molar mass of indium, and $M_{\rm Yb,Rh,Si}$ is the molar mass of Yb, Rh and Si in the given composition (e.g., for Yb:Rh:Si\,=\,3:2:2, $M_{\rm Yb,Rh,Si}= 781.1$\,g/mol). During variation of the stoichiometry, we kept all other growth parameters constant.

In Fig.~\ref{ratio}a, we present the influence of the Yb-content on the crystal thickness, $\overline d$, where we take an average value for the 10 largest crystals of each batch. The Yb-content could be varied in a broad range from  1:2:2 to 5:2:2 (Yb:Rh:Si). In all cases stoichiometric \YRSs single crystals are formed, however, the different Yb-content of the initial melt leads to single crystals with different thicknesses. From Fig.~\ref{ratio}a it is evident that the thickest crystals ($\overline d\approx 0.45$\,mm)  could be obtained with a three-times higher Yb-content, i.e. ($n_{\rm Yb}/n_{\rm Rh, Si}=1.5$). For higher Yb-contents, the thickness decreases and for $n_{\rm Yb}/n_{\rm Rh, Si}=2.5$ only very tiny single crystals are formed.

In a next step, we have optimized the ratio of Rh to Si, while keeping $n_{\rm Yb}/n_{\rm Rh, Si}=1.5$ constant. Here we learned, that the applicable range is much narrower, compared to the Yb one. The $n_{\rm Rh}/n_{\rm Si}$ ratio can only be varied from 0.9 to 1.35. In this range several crystal growths were carried out, with otherwise identical  parameters. We found that the mass of the crystals, $\overline m$, (again we take an average over the 10 largest crystals) is largest at $n_{\rm Rh}/n_{\rm Si}\approx 1.22$ corresponding to an initial ratio of Yb:Rh:Si\,=\,3:2.2:1.8 (Fig.~\ref{ratio}b, left axis). The thickness is nearly constant from $n_{\rm Rh}/n_{\rm Si}= 1$ to 1.22 ($\overline d\approx 0.5$\,mm) but considerably smaller ($\overline d\approx 0.3$\,mm) for $n_{\rm Rh}/n_{\rm Si} = 0.9$ and  $n_{\rm Rh}/n_{\rm Si} = 1.35$. More important is the influence on the crystal quality, reflected in the resistivity ratio at 1.8\,K, $\rho_{300\rm K}/\rho_{1.8\rm K}$, which increases from 5 for excess Si to 17.5 for $n_{\rm Rh}/n_{\rm Si}=1.35$ (Fig.~\ref{ratio}b, right axis). However, \YRSs is not anymore the main phase using this high Rh-content and YbRhIn$_5$ is formed instead. This is reflected in overall much smaller and thinner crystals for $n_{\rm Rh}/n_{\rm Si}=1.35$. Therefore, we decided to continue the optimization procedure of the temperature profile with the atomic ratio Yb:Rh:Si\,=\,3:2.2:1.8.

\begin{figure}[t]
\begin{center}
\includegraphics[width=0.8\textwidth]{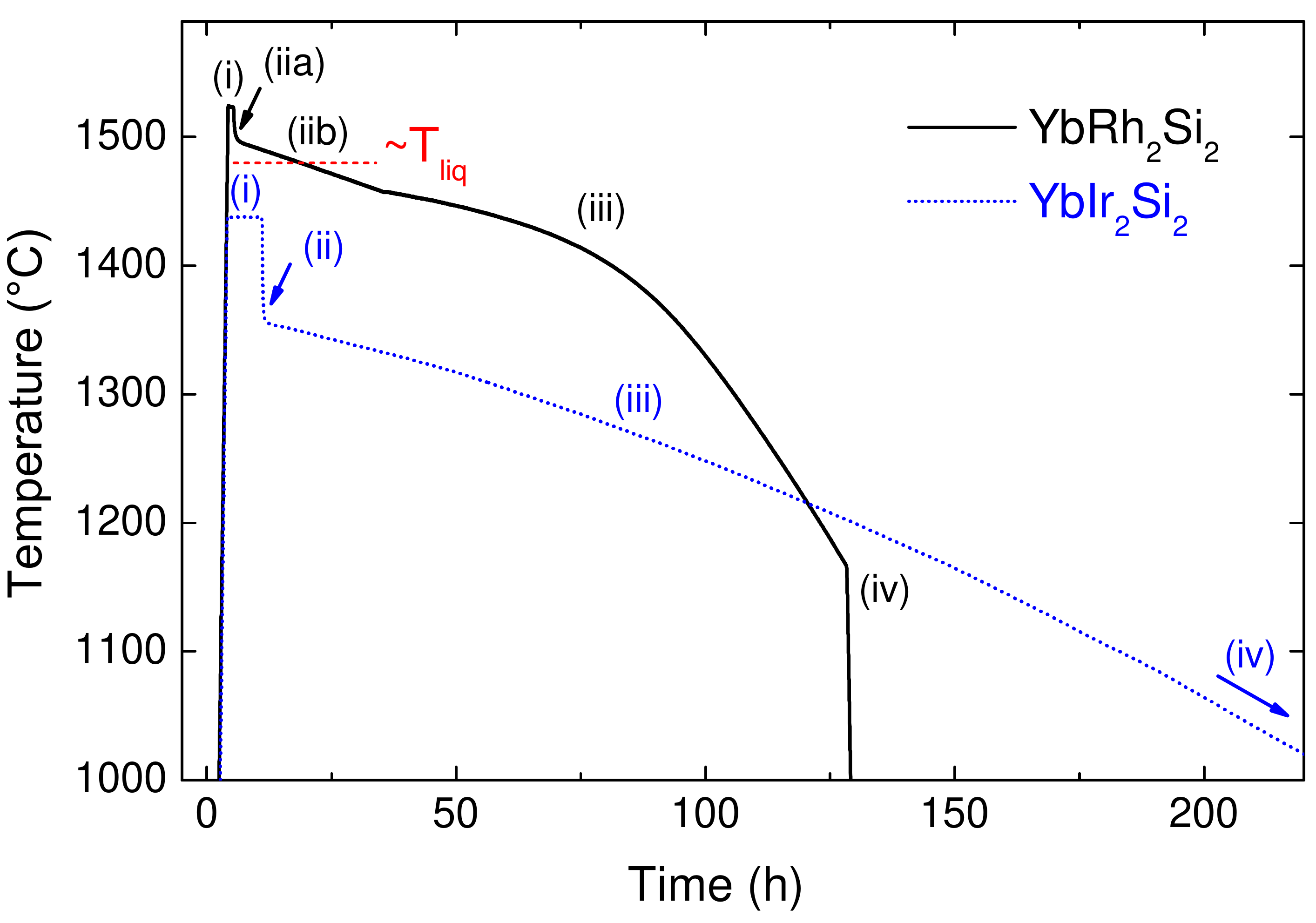}
\caption{\label{temp} 
Optimized temperature-time profile for the single crystal growth of \YRSs and \YIS. The temperature was measured in-situ at the bottom of the crucible.}%
\end{center}
\end{figure}

Differential-thermal-analysis measurements were carried out to determine the liquidus temperature of \YRSs in 96 mol\% In. We found an anomaly in the cooling curve at $T_{\rm liq}\approx 1480^{\circ}$C, which most likely is due to the formation of \YRSs within the liquid indium. From several growths, not following a systematic optimization procedure, we  observed  that a reduction of the time at highest temperatures results in higher-quality single crystals. Therefore, we have optimized the temperature-time profile, which includes several steps (solid line in Fig.~\ref{temp}), by using a vertical two-zone furnace (Xerion company).

First, we heated both zones  to reach a temperature of $1520^{\circ}$C throughout the crucible for one hour (step (i) in Fig.~\ref{temp}). Subsequently, a temperature gradient was build up within the crucible by moving the sample out of the furnace (step (iia)) and by slowly decreasing the power of the lower heating zone (step (iib)).  Finally, we start the Bridgman-type cooling by moving the Ta-tube further out of the furnace, whereas the higher heating zone is kept at its initial temperature (step (iii)). The crystal growth is finished at around $1150^{\circ}$C and the furnace is cooled down to room temperature (step (iv)). 

We assume that the following processes take place: Step (i) provides an homogenization of the melt  at temperatures well above $T_{\rm liq}$. The first seeds of \YRSs can form at the bottom of the crucible when the temperature goes through $T_{\rm liq}$ (step (iib)). We use the second heating zone to build up a temperature gradient inside the crucible, which significantly reduces the total time at high temperatures. The crystal grow during step (iii), where the temperature-time curve has at the beginning a much smaller slope. In contrast to a normal Bridgman technique without a metallic flux, here, several seeds form at the bottom of the crucible, which result in many crystals of comparable size and quality.

\begin{figure}[t]
\begin{center}
\includegraphics[width=0.8\textwidth]{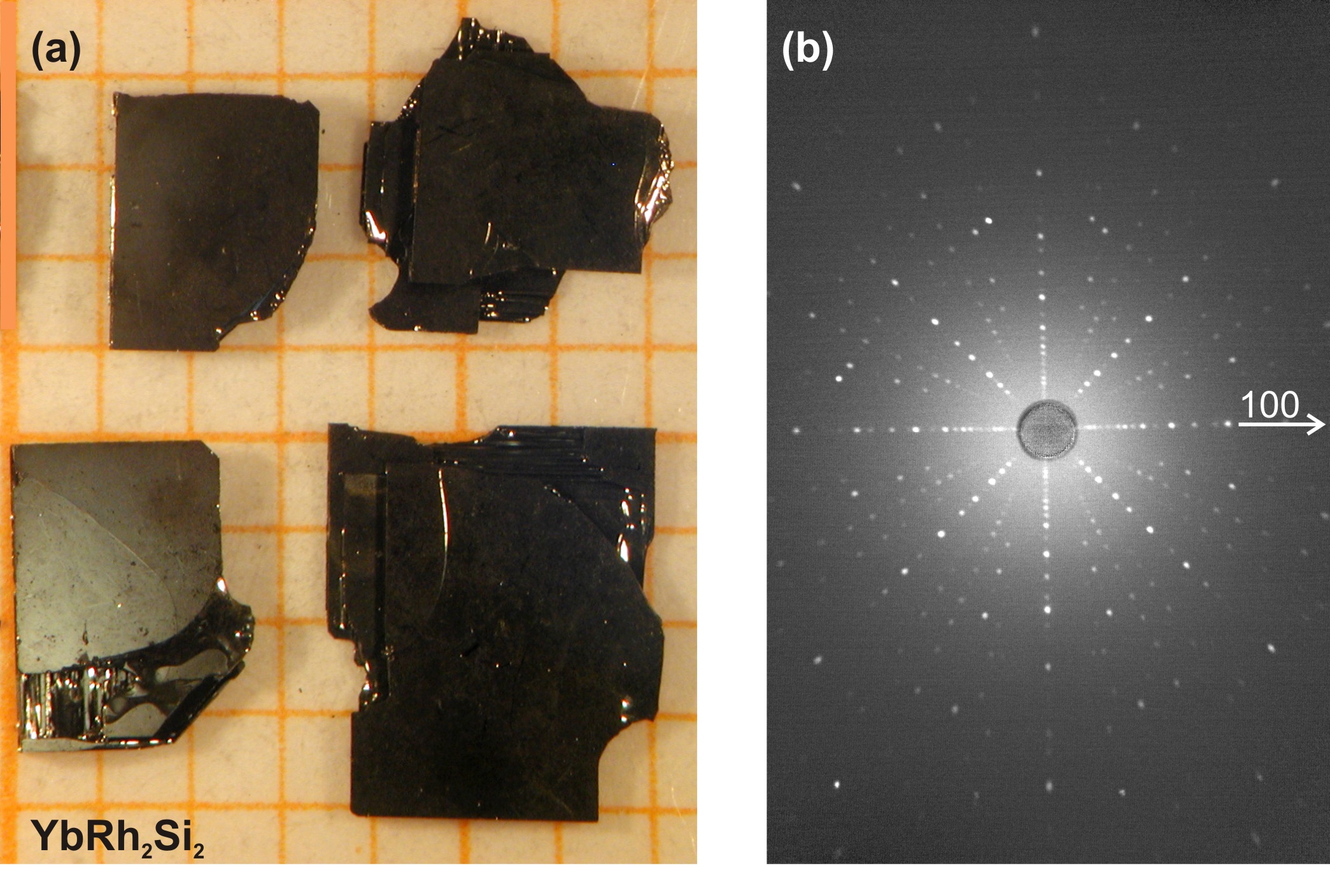}
\caption{\label{xtal} 
(a) Photograph of high-quality \YRSs single crystals. The edges are statistically aligned along 100 and 110, which can be proved by Laue back scattering images plotted in (b). Here, the horizontal direction corresponds to the 100 direction.}%
\end{center}
\end{figure}

The optimization of the temperature-time profile leads to single crystals with much higher resistivity ratios, reaching $\rho_{300\rm K}/\rho_{1.8\rm K}=27$. Some of these crystals are pictured in Fig.~\ref{xtal}a. These new generation of single crystals were deeply investigated at low temperatures and reveal a broad range of exciting physical phenomena. The residual resistivity, extracted from measurements down to the mK range, reaches $\rho_0\approx 0.5\,\mu\Omega$cm resulting in a resistivity ratio as high as $\rho_{300\rm K}/\rho_0\approx 150$ \cite{Gegenwart:2008a}. The extremely sharp anomaly in the specific heat at $T_N$ is a further proof of the high quality \cite{Krellner:2009}. Directly visible is the low defect density of these crystals in recent scanning tunneling measurements \cite{Ernst:2011}, where large areas of defect-free surfaces could be resolved. Some of the few defects could be assigned to Rh occupation on Si-sites, in agreement with the slight Rh excess described above. In contrast, single crystals prepared with excess Si show a lot more defects at the surface \cite{Wirth:2011}, which confirms that the resistivity ratio is a valid measure to access the quality of the crystals.

\subsection{\YIS}\label{SecYISgrowth}
The growth of the related compound \YISs is possible using a similar high-temperature In-flux method \cite{Hossain:2005}. Here, we used an one-zone resistance furnace and a slightly different temperature-time profile during the crystal growth (see dotted curve in Fig.~\ref{temp}). The homogenization (step (i)) is done for 8 hours at $1450^{\circ}$C. The temperature gradient was build up by moving the sample out of the furnace, which leads to a temperature jump of $100^{\circ}$C ((step (ii) in Fig.~\ref{temp}). Subsequently, we start the Bridgman-type cooling by slowly moving the sample tube further out of the furnace (step (iii)). After 300 hours, the crystal growth is terminated at around $800^{\circ}$C.

Using a stoichiometric composition of the elements with 96\,mol\% indium, gives quite large crystals with an average mass of 35\,mg and an average thickness of 0.6\,mm. Furthermore, the purest crystals obtained without optimization have already a residual resistivity of only $\rho_0\approx0.3\,\mu\Omega$cm giving a resistivity ratio, $\rho_{300\rm K}/\rho_0\approx 225$ \cite{Hossain:2005}. 

The reproducibility of the growth process is not as good as for \YRS. Nearly identical growth parameters lead sometimes to large but also to small crystals, making a quantitative optimization very difficult. However, we varied also for \YISs the initial composition and found that the applicable ranges are much smaller than for \YRS. Doubling the Yb-content, $n_{\rm Yb}/n_{\rm Ir, Si}=1$, do not yield \YISs crystals anymore, instead binary phases as e.g., YbIn$_2$, YbIr$_3$ and Yb$_5$Si$_3$ are formed. A small excess of Yb, $n_{\rm Yb}/n_{\rm Ir, Si}=0.6$,  leads to the formation of P-type \YIS. Using an off-stoichiometric ratio of Ir to Si, $n_{\rm Ir}/n_{\rm Si}= 1.22$, do no result in \YISs crystals but to the formation of some unknown phases. From  these results it is obvious that the optimal crystal growth parameters are much different in \YRSs and \YIS. However, the present output, concerning size and quality of the \YISs crystals is already satisfactory, so no further optimization was undertaken.

The formation of the two structural variants, I-type and P-type, in \YISs is not surprising, because these two different structures were observed in other RIr$_2$Si$_2$ compounds as well \cite{Braun:1983, Buffat:1986, Niepmann:2001, Mihalik:2011}
. There, it was established, that the P-type configuration forms only at high temperatures, whereas the I-type is stabilized at lower temperatures. Here, we are mainly interested in the I-type structure which is isostructural to \YRS. Therefore, we have chosen the temperature-time profile such that a long annealing time below $1000^{\circ}$C was achieved (see Fig.~\ref{temp}). It is presently unclear, why the excess of Yb leads to the stabilization of the P-type phase. However, we were able to  transform the P-type single crystals into I-type by annealing the crystals  at $1000^{\circ}$C for 150 hours. For this purpose the crystals extracted from the In-flux were enclosed in a Ta-crucible under argon at ambient pressure. We discuss the differences of I- and P-type in the X-ray diffraction patterns and the different physical ground states in the sections \ref{SecXrayPI} and \ref{SecPI}, respectively.

\subsection{\YRIS}
As stated above, negative chemical pressure has to be applied to \YRSs to reach the QCP in the absence of external magnetic field. This can be achieved by various kind of substitutions. Ge-substitution on the Si site was tried first \cite{Trovarelli:2002, Custers:2003}, however, the maximal substitution level amounts to only about 2\,at\% \cite{Ferstl:2007}. Higher substitution can be achieved replacing the magnetic Yb-atoms with non-magnetic La \cite{Ferstl:2005} or Lu \cite{Koehler:2008a}. Here, we discuss a further substitution series, \YRIS, where the Rh atoms are replaced by the larger isoelectronic Ir-atoms.

\begin{table}[tb]
\tbl{Parameters and results of the crystal growth of \YRIS.}
{\begin{tabular}{@{}lccccc}\toprule
Initial composition& $x_{\rm Ir}^{\rm nom}$ & $x_{\rm Ir}^{\rm EDAX}$ & $\overline d$ & $\overline m$ & $\frac{\rho_{300\rm K}}{\rho_{1.8\rm K}}$  \\
 $\rm Yb:Rh:Ir:Si$ 	& &  & (mm) & (mg) &	 \\ \hline  
$3:1.96:0.04:2.0$ & 0.02	& 0.025	 & 0.4	& 13 & 7.1	\\ 
$3:1.90:0.10:2.0$ & 0.05	& 0.06		 & 0.2  & 4  & 6.5 \\ 
$3:1.60:0.40:2.0$ & 0.2		& 0.17	 & 0.3	& 4  & 3.9 \\  
$3:1.68:0.42:1.9$ & 0.2 	& 0.23	 & 0.2  & 6  & 4.1 \\ 
\botrule
\end{tabular}}
\label{TabIrdop}
\end{table}

For the crystal growth of \YRIS\, we used the optimized growth parameters of \YRS, with the respective replacement of Rh by Ir. The initial stoichiometry together with the properties of the resulting single crystals are given in Table~\ref{TabIrdop}. For all growths the amount of In-flux was kept constant as described above, with $m_{\rm In}/m_{\rm Yb,Rh,Ir,Si}= 24 M_{\rm In}/M_{\rm Yb,Rh,Ir,Si}$. 

The Ir content was determined using energy dispersive X-ray (EDAX) spectra obtained in a scanning electron microscope (Philips XL30) with a Si(Li)-X-ray detector. Three crystals of each batch with a polished surface were analyzed at several positions. The respective mean values result in the given $x_{\rm Ir}^{\rm EDAX}$ values with an absolute error of less than $0.01$. These real Ir-concentrations, which are used for $x_{\rm Ir}$ throughout this article, are somewhat different than the nominal concentrations, $x_{\rm Ir}^{\rm nom}$,  due to different solubilities of the elements in the In-flux. The average thickness and mass of the single crystals are again determined by averaging over the 10 largest crystals. The resistivity ratio, $\rho_{300\rm K}/\rho_{1.8\rm K}$, reflects the amount of disorder in the single crystals.

Unfortunately, the size of the crystals gets smaller with increasing Ir-content, opposite to what was observed for the \YRCS\, series, where the crystal growth gives  larger crystals, with increasing $x_{\rm Co}$ \cite{Klingner:2011}. The value of the resistivity ratio decreases with increasing $x_{\rm Ir}$, which is expected due to the higher amount of disorder introduced by the statistical occupation of the Ir atoms at the Rh site. An exception is the growth with $x_{\rm Ir}=0.23$, because  $\rho_{300\rm K}/\rho_{1.8\rm K}$ is somewhat larger than for $x_{\rm Ir}=0.17$. This is most probably caused by the higher Rh/Ir-content with respect to Si ($n_{\rm Rh+Ir}/n_{\rm Si}=1.1$), similar to what was observed in the pure \YRS, where we found a larger resistivity ratio for higher Rh-content (see Fig.~\ref{ratio}).

\section{X-ray diffraction}
\subsection{\YRS}\label{SecLaue}
A series of 10 samples of different quality has been systematically analyzed by X-ray diffraction and electron microprobe analysis. The lattice parameters were accurately determined in respect to the chemical composition and the homogeneity range of the phase \YRSs was defined. These results and the experimental details will be published elsewhere \cite{Wirth:2011}.

In Fig.~\ref{xtal}b, we present a X-ray Laue backscattering image from a \YRSs single crystal. The reflection peaks are pin sharp rather than blurred, proving again the high quality of the synthesized crystal. These pictures were used to orient the single crystals. In all cases, the direction perpendicular to the surface corresponds to the crystallographic $c$-axis, proved by the fact that the 001 reflection is exactly in the middle of the aperture. Simulations demonstrate that the horizontal direction corresponds to the 100 direction, which however, not always corresponds to the edges of the crystal, visible in Fig.~\ref{xtal}a. Our experience from the orientation of more than 100 of these crystals for photo-emission spectroscopy, revealed that the crystal edges much more often corresponds to the 110 direction, although not exclusively. 

\subsection{\YIS}\label{SecXrayPI}
\begin{figure}[t]
\begin{center}
\includegraphics[width=0.8\textwidth]{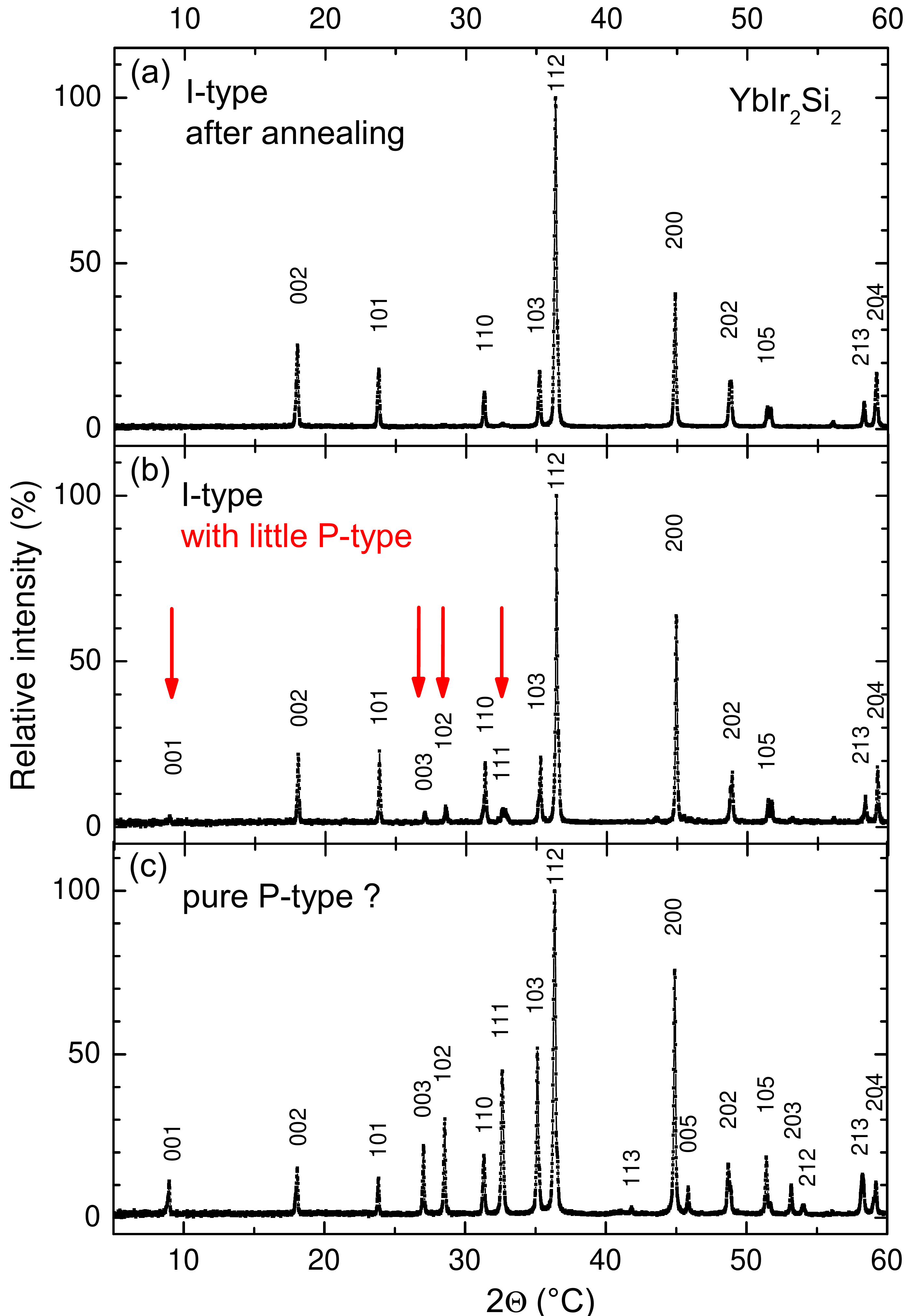}
\caption{\label{fig5xrd} 
Comparison of three different X-ray powder diffraction patterns of \YIS. (a) I-type after annealing. (b) I-type with small amount of P-type, leading to additional peaks at  $hkl=001,\, 003,\, 102,\, 111$ (arrows). (c) P-type pattern; the intensities of the peaks have changed considerably and all odd $hkl$ are observed. However, small amount of I-type within the P-type phase cannot be excluded.}%
\end{center}
\end{figure}

The distinction between the body-centred and the primitive structure type of \YISs is possible using powder X-ray diffraction, which was performed on a Stoe diffractometer in transmission mode using monochromated Cu-K$_{\alpha}$ radiation ($\lambda = 1.5406$\,\AA). For the primitive structure additional peaks ($hkl=001,\,003,\,102,\,111$) are observable. For the body-centred structure these peaks with an odd sum of $hkl$ are absent. 

In Fig.~\ref{fig5xrd} we present three different diffractograms for a) pure I-type \YIS, b) I-type with small content of P-type and c) mainly P-type. The diffraction pattern of the annealed single crystals only show very tiny odd $hkl$ peaks, confirming that in this case we have virtually pure I-type crystals. When the crystal growth is terminated before \YISs is completely transformed into the I-type structure, the diffraction pattern presents small peaks at the $hkl=001,\,003,\,102,\,111$ positions (arrows in Fig.~\ref{fig5xrd}b). The growth with Yb-excess leads to P-type \YIS, the corresponding pattern is shown in Fig.~\ref{fig5xrd}c. Here, the intensities of the peaks have changed considerably and all odd $hkl$ peaks are observed. However with powder diffraction, we cannot finally prove, whether parts of the P-type crystal are already transformed into I-type. This is because the lattice parameters of the two structures are nearly identical with $a_{\rm I}=4.0345(15)$\,\AA, $a_{\rm P}=4.0370(20)$\,\AA\, and $c_{\rm I}=9.828(2)$\,\AA, $c_{\rm P}=9.898(5)$\,\AA, which were refined by simple least squares fitting. 

\subsection{\YRIS}\label{Seclattice}
\begin{figure}[t]
\begin{center}
\includegraphics[width=0.8\textwidth]{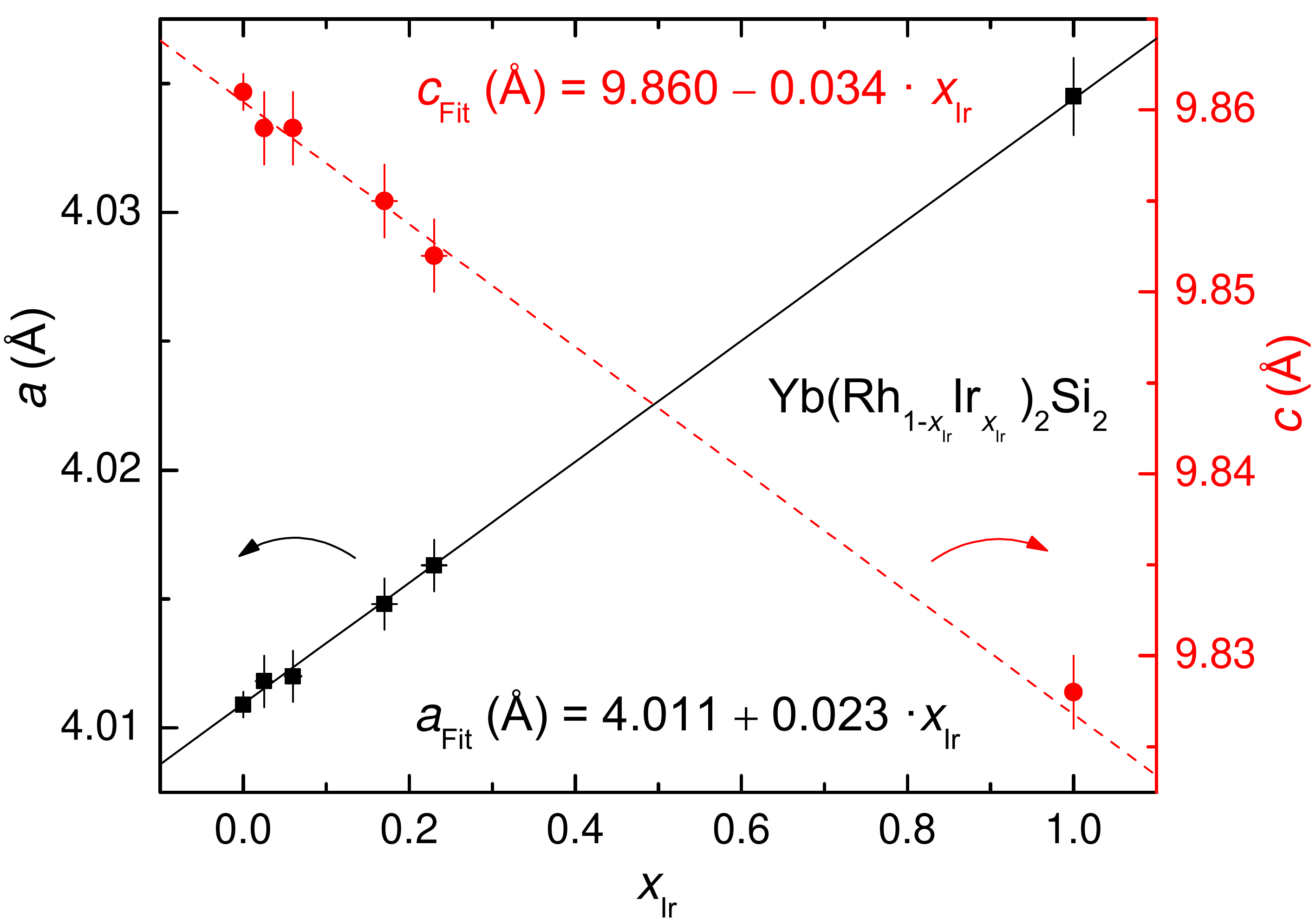}
\caption{\label{fig6ac} 
Lattice parameters $a$ (left axis) and $c$ (right axis) for the \YRIS\, series. With increasing Ir content, $a$ linearly increases, whereas $c$ decreases.}%
\end{center}
\end{figure}

In Fig.~\ref{fig6ac} the lattice parameters $a$ (left axis) and $c$ (right axis) are plotted for the different Ir-concentrations. An important point is immediately visible, namely that $a$ and $c$ have opposite trends with increasing $x_{\rm Ir}$. The lattice parameter $a$, which describes the distance of the Yb-ions in the plane (see Fig.~\ref{fig1}), increases linearly with $x_{\rm Ir}$ and can be described with $a_{\rm Fit}$(\AA) $=4.011+0.023\cdot x_{\rm Ir}$. In contrast, $c$ decreases linearly, with $c_{\rm Fit}$(\AA) $=9.860-0.034\cdot x_{\rm Ir}$. The resulting volume of the unit cell increases with increasing $x_{\rm Ir}$, however, the overall volume change going from \YRSs to \YISs amounts to only $\Delta V = 1.3(2)$\,\AA$^3$, which is less than 1\%. This is a rather small increase of the unit cell volume, because the variation of $a$ and $c$ cancels out each other. The corresponding hydrostatic pressure, if applied to \YIS, amounts to $p\approx 1.6$\,GPa, using a bulk modulus of 190\,GPa \cite{Plessel:2003}. However, for increasing Ir-concentration, the ratio of the lattice parameters, $c/a$, strongly decreases, as $c$ decreases and $a$ increases with $x_{\rm Ir}$. This results in a distortion of the unit cell, which gets less elongated with higher Ir-content. Therefore, isoelectronic substitution is not equivalent to hydrostatic pressure, which would result in a constant $c/a$ ratio. This is probably the reason why for \YISs a huge pressure of 8\,GPa is needed to induce magnetic order \cite{Yuan:2006}, in contrast to the expected 1.6\,GPa mentioned above. However, for small $x_{\rm Ir}$ the relative change of $c/a$ is less than 0.1\%, compared to pure \YRS, so that small Ir-substitution is equivalent to hydrostatic pressure \cite{Macovei:2008, Friedemann:2009a}.


\section{Physical characterization}
\subsection{Temperature exponent of the electrical resistivity in \YRS}\label{Secexp}
\begin{figure}[t]
\begin{center}
\includegraphics[width=0.8\textwidth]{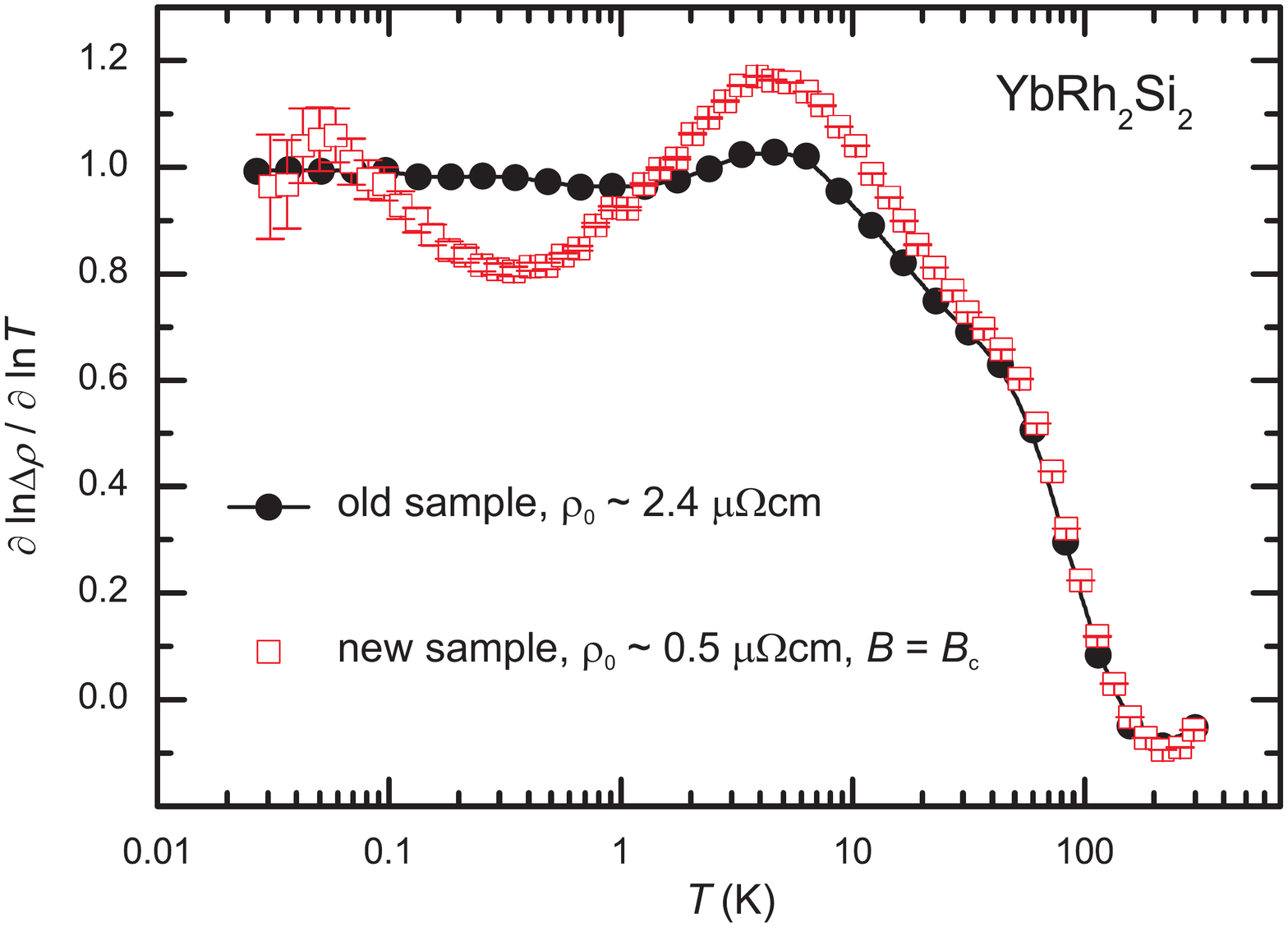}
\caption{\label{fig7exp} 
Temperature dependence of the electrical-resistivity exponent, $\varepsilon=\frac{\partial\ln\Delta\rho(T)}{\partial\ln T}$, for two  single crystals with different residual resistivity. With higher sample quality (lower $\rho_0$) pronounced deviations from the linear-in-$T$ resistivity are evident below 10\,K.}%
\end{center}
\end{figure}

One of the hallmarks of non-Fermi-liquid behaviour in \YRSs is the linear-in-$T$ resistivity over more than an order of magnitude in temperature. However, this was studied in a wider temperature range only for the very first generation of single crystals with $\rho_0\approx2.4\,\mu\Omega$cm  \cite{Trovarelli:2000a}. Here, we present an analysis of the temperature exponent of the resistivity for one of the optimized single crystals with a residual resistivity as low as $\rho_0\approx0.5\,\mu\Omega$cm. In zero magnetic field, the resistivity shows a sharp drop at the AFM phase transition ($T_N=70$\,mK) but a small critical magnetic field ($B_c=50\,$mT, $B\parallel j \perp c$) can be applied to tune the system towards quantum criticality, where the non-Fermi-liquid behaviour is observed down to lowest temperatures \cite{Gegenwart:2008a}.

Generally, the temperature dependence of the resistivity can be described with $\Delta\rho(T)=\rho(T)-\rho_0 \propto T^{\varepsilon}$. Therefore, the temperature exponent, $\varepsilon$, can be visualized by plotting $\partial\ln \Delta\rho(T) / \partial\ln T$, shown in Fig.~\ref{fig7exp} for two samples with different residual resistivity. The data for the old sample presents a nearly constant $\varepsilon=1$ below 7\,K, down to lowest measured temperatures. In contrast, the exponent of the new sample exhibits a pronounced maximum at around 4\,K with $\varepsilon\approx1.2$, followed by a minimum around 0.3\,K with $\varepsilon=0.8$ and a value of $\varepsilon=1$ at lowest temperatures. For this analysis we estimate $\rho_0$ by linear extrapolation to $T=0$, however, the choice of $\rho_0$ influences $\varepsilon(T)$ only at lowest temperatures. This is visible by the larger error bars of $\varepsilon$ below 0.1\,K, which result from the uncertainty of $\rho_0$. Beside this temperature-dependent analysis of the exponent, we additionally fit the data with a single $\varepsilon$ in a larger temperature range. For the resistivity data below 0.1\,K, we obtain an exponent of $\varepsilon=1.00\pm0.01$. A fit below 0.2\,K down to lowest temperatures gives  $\varepsilon=0.831\pm0.004$, in agreement with the differential analysis.

We note that the evolution of the exponent for these two different samples presents qualitative similarities to quasi-classical calculations for a system in the vicinity of an antiferromagnetic QCP. There, A. Rosch studied the influence of strong anisotropic scattering due to spin fluctuations for different amount of disorder and he observed broad ranges of exponents between 1 and 2 together with oscillatory behaviour \cite{Rosch:1999}. It might be interesting to implement the specifics of \YRSs to such a model to get further insight into the transport phenomena at this local QCP.


\subsection{Different ground states in I-type and P-type \YIS}\label{SecPI}

\begin{figure}[t]
\begin{center}
\includegraphics[width=0.8\textwidth]{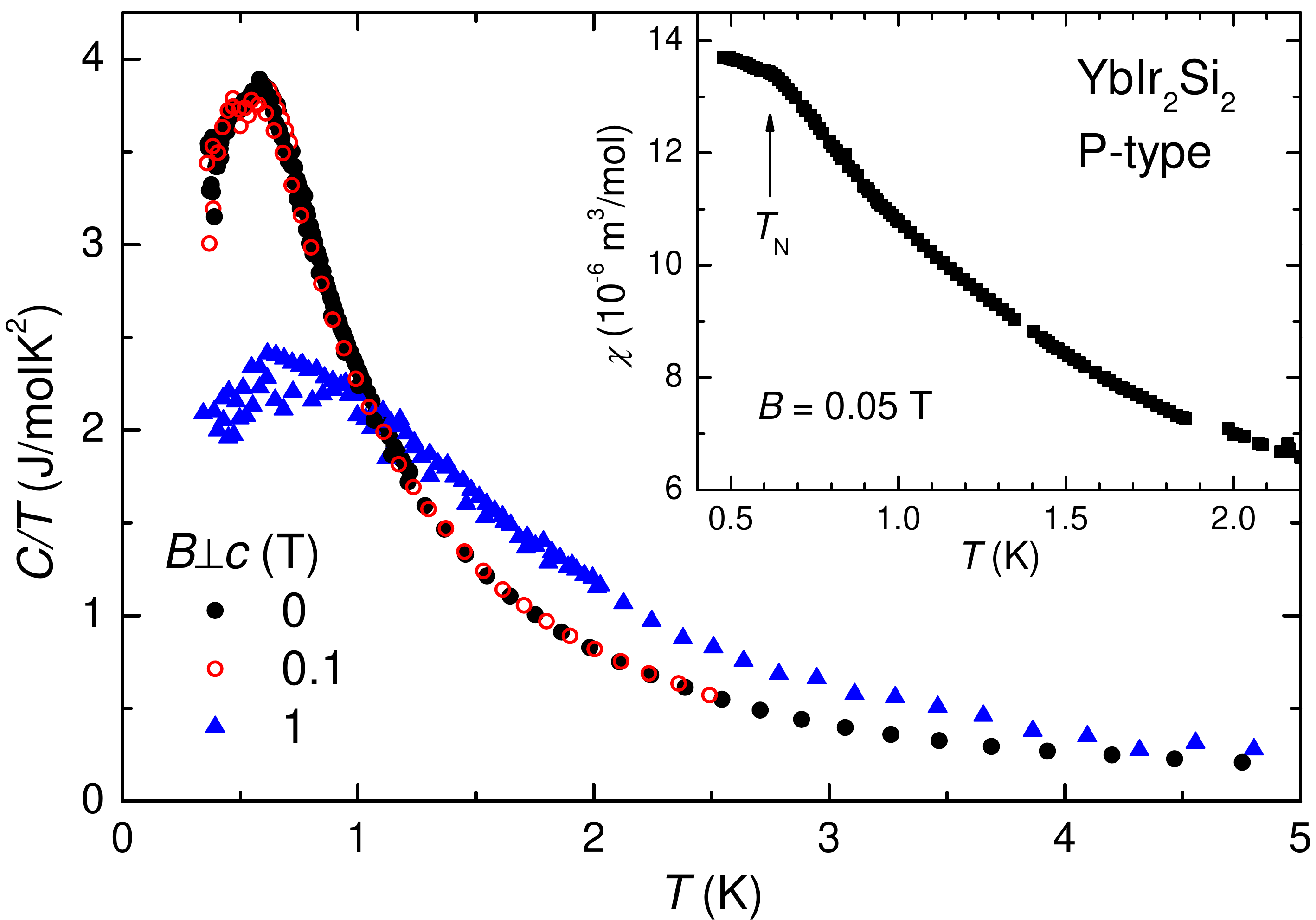}
\caption{\label{fig9ptype} 
Specific heat divided by temperature, $C/T$, of \YISs (P-type) on a linear temperature scale. At $T_{N}\approx 0.6$\,K, a pronounced anomaly is visible which is suppressed in $B=1$\,T. The dc susceptibility, plotted in the inset, confirms the magnetic nature of this phase transition.}%
\end{center}
\end{figure}

\YISs crystallizes in two different structure types, a bodycentred (I-type) and a primitive (P-type) variant (see Fig.~\ref{fig1} and Sec.~\ref{SecYISgrowth}). Interestingly, the magnetic ground states of these two derivatives of the BaAl$_4$ structure are very different, although, the tetragonal lattice parameters are nearly identical, as discussed in Sec.~\ref{SecXrayPI}. Hossain \EA\, have shown that I-type \YISs is a heavy-fermion metal situated on the non-magnetic side of a QCP with Fermi-liquid behaviour below 0.4\,K  and quite large single-ion anisotropy, reflected in a  larger susceptibility perpendicular to the crystallographic $c$-direction \cite{Hossain:2005}. On the other hand, the susceptibility of P-type \YISs is rather isotropic \cite{Hossain:2006} and entropy as well as resistivity measurements suggest a much smaller Kondo-temperature, $T_K^P\approx 2$\,K, in contrast to $T_K^I\approx 40$\,K for the I-type system \cite{Hossain:2005}. Furthermore, in P-type crystals a prominent anomaly was observed  in the specific heat below  1\,K, which was suggested to be a magnetic phase transition. 

Here, we further like to establish that P-type \YISs indeed orders antiferromagnetically at $T_N=0.6\,$K by means of susceptibility measurements below 2\,K. For this purpose, P-type crystals were prepared with Yb-excess as described in Sec.~\ref{SecYISgrowth} and the dc susceptibility was measured down to $T=0.5$\,K, which was realized with the $^3$He-option for the MPMS (Quantum Design) designed by the IQUANTUM Corporation. In the inset of Fig.~\ref{fig9ptype}, we present this measurement with $B=0.05$\,T perpendicular to $c$. A weak anomaly suggests that P-type \YISs undergoes a magnetic phase transition at $T_N\approx 0.6$\,K, although the transition is not very sharp. Specific heat measurements on these crystals confirm the earlier measurements in zero field \cite{Hossain:2005} and are shown in the main part of Fig.~\ref{fig9ptype}. Applying a magnetic field of 0.1\,T does not affect the anomaly at $T_N$, whereas a field of $B=1\,$T suppresses the magnetic transition and shifts the entropy to higher temperatures. The characteristics of this field dependence confirms the AFM nature of the phase transition and classifies P-type \YISs as a magnetically ordered system with localized Yb$^{3+}$ moments. However, the anomalies at $T_N$ are rather broad and we were not able to detect the magnetic transition in low-temperature resistivity measurements, probably due to significant amount of disorder, likely Ir-Si site exchanges. 

\section{Conclusion}
In conclusion, we have optimized the single crystal growth of \YRS, using an indium-flux method with 96 mol\% indium in closed Ta-crucibles. The optimal initial composition of the elements within the flux, Yb:Rh:Si\,=\,3:2.2:1.8, is far away from the stoichiometric ratio but was found to give large (up to 100\,mg) single crystals. Prior to the optimization of the temperature-time profile, we determined the liquidus temperature ($T_{\rm liq}\approx1480^{\circ}$C) of \YRSs in 96 mol\% In-flux by means of differential thermal analysis and use a two-zone resistance furnace to reduce the time of the crystal growth. The resulting single crystals have a residual resistivity of only $\rho_0\approx0.5\,\mu\Omega$cm and present sharp anomalies in the transport and thermodynamic properties at the low-lying AFM phase transition. The analysis of the temperature-dependence of the electrical-resistivity exponent for this new generation of single crystals reveals a deviation from the linear-in-$T$ behaviour below 10\,K.

\YISs was established to be a polymorphic compound which can crystallize in two different crystallographic modifications, with P-type crystals (\CBG\, structure) being the high-temperature and I-type (\TCS\, structure) the low-temperature phase. The latter can be stabilized by annealing selected crystals at $1000^{\circ}$C for 150 hours. We further found that the optimal growth parameters  for \YRSs and \YISs from In-flux are much different. For \YIS, the applicable range of the initial composition is very narrow and small excess of Yb leads to the formation of P-type crystals. Interestingly, both modifications have different physical ground states. In P-type \YIS, the magnetic $4f$-electrons undergo an AFM phase transition at $T_N\approx0.6$\,K, observed in low-temperature specific-heat and susceptibility measurements, whereas I-type \YISs presents a paramagnetic Fermi-liquid ground state and is situated on the non-magnetic side of the quantum critical point. The series \YRIS, which crystallizes in the I-type structure, is therefore ideally suited to study the phenomena around this quantum phase transition.

\section*{Acknowledgements}
The authors thank U. Burkhardt and P. Scheppan for energy dispersive X-ray analysis of the samples as well as N. Caroca-Canales and R. Weise for technical assistance.
The presented low-temperature susceptibility measurements were performed with the help of C. Klausnitzer.
We acknowledge valuable discussions with
M. Brando,
M. Deppe,
J. Ferstl,
S. Friedemann, 
P. Gegenwart,
A. Rosch,
and
F. Steglich.
The DFG (Research Unit 960, "Quantum phase transitions") is acknowledged for financial support.

\end{document}